\documentclass[]{spie}  

\usepackage{amsmath,amsfonts,amssymb}
\usepackage{graphicx}
\usepackage[colorlinks=true, allcolors=blue]{hyperref}

\title{Quantum control of solid-state qubits \\ for thermodynamic applications}

\author{Paul R. Eastham}
\author{Conor Murphy}
\affil{School of Physics, Trinity College Dublin, Dublin 2, Ireland}

\DeclareMathOperator{\Tr}{Tr}

\begin{document} 
\maketitle

\begin{abstract}
We give an overview of our recent theoretical studies of the thermodynamics of excitons, and other solid-state qubits, driven by time-dependent laser fields. We consider a single such emitter and describe how the formation of strong-field dressed states allows the emitter to absorb or emit acoustic phonons in a controlled way. We present results for the heat absorption, and show that the form of the driving field can be tailored to produce different thermodynamic processes, including both reversible and irreversible heat absorption. We discuss these effects from the perspective of quantum thermodynamics and outline the possibility of using them for optical cooling of solids to low temperatures. 
\end{abstract}


\section{INTRODUCTION}
\label{sec:intro}  

Quantum control\cite{ramsay_review_2010} uses tailored driving of a quantum system to achieve a targeted dynamical process or state. The simplest form uses an applied field, which is typically a tailored laser or microwave pulse, to drive a quantum system into a specific target state. For open quantum systems, which interact with their environments, there can be other desired outcomes of the control. One can seek driving fields, such as the Hahn echo sequence, which recover information dissipated to an environment to the system, or minimize its decoherence. A different perspective appears if we consider the effects of the control on the environment as well as the system: can one use a particular form of driving to produce a given effect, such as cooling, in the environment? This is, of course, what occurs in laser cooling. More generally, however, an understanding of heat flows and their control in quantum systems is relevant for the design and optimization of quantum thermal machines, such as lasers, photovoltaics, and even clocks.

In this contribution, we present some of our theoretical results on
the thermodynamics of driven solid-state quantum systems. We focus on
a quantum-dot exciton, which is, in some circumstances, a two-level
system that can be controlled by driving with ultrafast laser
pulses. Such pulses have been used to create a single-exciton state,
by using Rabi flopping\cite{zrenner_coherent_2002}, with resonant
excitation, and by using adiabatic rapid
passage\cite{wu_population_2011,simon_robust_2011,malinovsky_general_2001}, with chirped
excitation. Phonons in the host semiconductor have been shown to
inhibit the creation of an exciton using these particular
protocols\cite{luker_influence_2012,eastham_lindblad_2013,ramsay_damping_2010,forstner_phonon-assisted_2003,machnikowski_resonant_2004,nazir_modelling_2016}. However,
other protocols can be used to create an exciton with the assistance
of phonons\cite{glassl_proposed_2013,quilter_phonon-assisted_2015}.

Here, we discuss the exciton-phonon dynamics in a driven quantum dot
from the perspective of thermodynamics, and report forms of driving
pulse which enable heat to be transferred from the phonons to the
exciton. We summarize results from our recent paper\
\cite{murphy_quantum_2019} for the heat absorption of different
pulses, and also the associated entropy generation, which quantifies
the extent to which the heat transfer is reversible. These results
show that suitable shaping of the frequency profile of the driving allows
for near-reversible heat transfers, in effect implementing an
isothermal expansion or compression. The heat distributions are
evaluated, and found to be strongly non-Gaussian, and reflective of
the form of the driving. We present a preliminary analysis of a
steady-state cooling process based on phonon-assisted transitions
between laser-dressed states. While for excitons the inhomogeneous
broadening precludes effective cooling, this analysis suggests that
other solid-state systems could be cooled to low temperatures using this method.

\section{Modelling Exciton Dynamics and Thermodynamics}
\label{sec:modelling}

\begin{figure}
    \centering
    \includegraphics{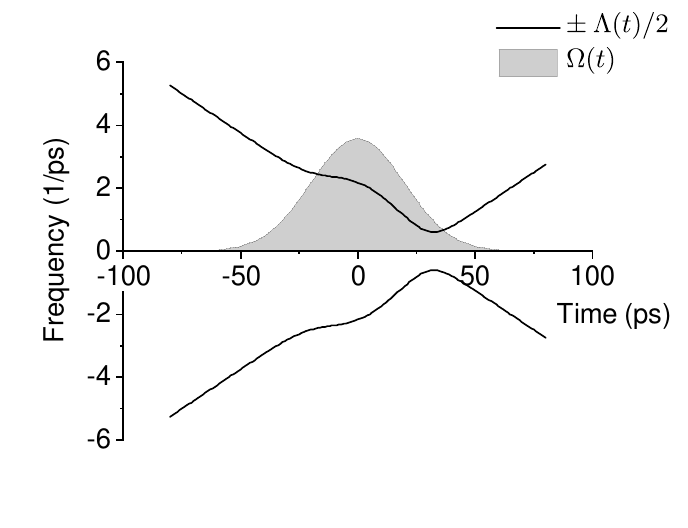}
    \caption{Dressed-state energy levels, $\pm\Lambda/2$, and Rabi frequency, $\Omega(t)$, for a two-level system driven with a chirped laser pulse.}
    \label{fig:carnotpulselevels}
\end{figure}

There are many important examples of few-level quantum systems, in
solid-state environments, that can be driven using electromagnetic
fields. We consider the example of an exciton -- a bound electron-hole
pair -- confined to an InGaAs/GaAs quantum dot. For simplicity, we
model the dot as a two-level system, formed from the ground state
$|0\rangle$ and a single one-exciton state $|1\rangle$. This model
corresponds directly to the case of driving near resonance with
circularly polarized light, as we are neglecting both the excited
states and the exciton spin\cite{schmidgall_population_2010}. The
Hamiltonian describing the driven transition is then $ \hat{H}_{s}=E_x\hat{s}_z-d E(t) \hat{s}_x,$ where $E_x$ is the transition
energy, $d$ the transition dipole moment, and $E(t)$ the electric
field of the laser pulse. The pulse is expressed in terms of its
amplitude and frequency as
$E(t)=\tilde{E}(t)\cos(\int^t \omega(\tau) d\tau)$. Note that we use
units such that $\hbar=1$
throughout.
$\hat{s}_z=\frac{1}{2}(|1\rangle\langle 1|-|0\rangle\langle 0|)$ and
$\hat{s}_x=\frac{1}{2}(|1\rangle\langle 0|-|0\rangle\langle 1|)$ are
the standard pseudospin representation for the two-level system. After
making the time-dependent unitary transformation
$\hat{U}=e^{-i\omega(t)\hat{s}_z}$, and the rotating-wave
approximation, the Hamiltonian for the driven transition becomes \begin{equation}
\hat{H}_s=\Delta(t) \hat{s}_z-\Omega(t)\hat{s}_x, \label{eq:tls} \end{equation} where
$\Delta(t)=E_x-\omega(t)$ is the transition frequency relative to that
of the laser, and $\Omega(t)=d \tilde{E}(t)$ is the time-dependent Rabi frequency
associated with the driving.

The strong driving of the transition produces laser-dressed states,
$|+\rangle$ and $|-\rangle$, which are superpositions of the zero and
one-exciton states. The time-dependent energies of these dressed
states can be found straightforwardly by diagonalizing $\hat{H}_s$,
and are $\pm \Lambda/2$, where $\Lambda=\sqrt{\Delta^2+\Omega^2}$. An
example is shown in Fig.\ \ref{fig:carnotpulselevels}, for driving with a linearly-chirped
Gaussian pulse. The anticrossing between the dressed states, visible
in this figure, enables such pulses to be used to create a single
exciton in an adiabatic rapid passage
process\cite{wu_population_2011,simon_robust_2011}. This works because
the initial state, before the laser pulse, is the $|0\rangle$ state,
which, for these parameters, coincides with the lower-energy dressed
state, $|-\rangle$. After the pulse has passed, however, this lower-energy dressed state corresponds to the $|1\rangle$ state. If the time variation of the Hamiltonian, through
the parameters $\Delta(t)$ and $\Omega(t)$, is slow compared with the
inverse of the gap then the system will remain in this dressed state,
and so evolve from the zero to the one exciton state.

A complete description of the exciton dynamics needs to take into
account the coupling of the quantum-dot exciton to its environment:
the dot is an open quantum system rather than an isolated one. The
coupling to the electromagnetic environment leads to radiative decay
of the exciton, with a typical lifetime on the order of
nanoseconds. In addition, under strong driving there are significant
affects arising from the coupling between the excitons and the phonons
in the host crystal. The impact of this coupling on quatum control has
been modelled using a variety of
approaches\cite{luker_influence_2012,eastham_lindblad_2013,ramsay_damping_2010,forstner_phonon-assisted_2003,machnikowski_resonant_2004,nazir_modelling_2016}. Our
analysis is based on the methodology described in Ref.\
\citenum{eastham_lindblad_2013}, which gives rise to a master equation
of Lindblad
form, \begin{equation}\frac{d\hat{\rho}}{dt}=-i[\hat{H},\hat\rho]-\gamma_{a}\mathcal{L}(|+\rangle\langle-|)\hat{\rho}-\gamma_{e}\mathcal{L}(|-\rangle\langle+|)\hat{\rho}.\label{eq:diss} \end{equation}
Here $\hat{\rho}$ is the reduced density matrix of the dot, obtained
by tracing out the phonon environment. The first term is the coherent
dynamics of the exciton. The second and third are incoherent processes due
to the phonons. We use $\mathcal{L}(\hat{O})$ to denote the Lindblad
superoperator for a jump operator $\hat{O}$, whose action on the
density matrix is \begin{equation}
  \mathcal{L}(\hat{O})\hat{\rho}=\hat{O}^\dagger
  \hat{O}\hat{\rho}+\hat{\rho}\hat{O}^\dagger
  \hat{O}-2\hat{O}\hat{\rho}\hat{O}^\dagger.\label{eq:linsupop}\end{equation}
Since Eq. (\ref{eq:diss}) is in a standard Lindblad form, its physical
interpretation is straightforward: phonon absorption causes
transitions from the lower-energy dressed state, $|-\rangle$, to the
upper-energy one, with rate $\gamma_a$, and phonon emission causes
transitions in the opposite direction, with rate
$\gamma_e$. The absorption rate is $\pi\Omega^2n_B(\Lambda)J(\Lambda)/(2\Lambda^2)$, where $J(\Lambda)$ is the phonon spectral density, and $n_B(\Lambda)$ the phonon occupation, at energy $\Lambda$. The emission rate is similar, with a factor of $n_B+1$ instead of $n_B$. It will turn out that the thermodynamic effect of the emission and absorption processes is, as one might expect, to transfer an amount of heat $\Lambda$ between the dot and the phonon environment.

This methodology can be generalized to calculate thermodynamic
quatities, such as heat and work, using a counting field, or phase
marker, approach\cite{esposito_nonequilibrium_2009}. The heat
transferred between two times, $Q$, is the difference between the energies of the heat bath at those times (the work, meanwhile, is the difference in
dot energies). This is, in quantum thermodynamics, a stochastic
quantity, with probability distribution $P(Q)$. It can be calculated
from the two-time correlation function of the bath Hamiltonian,
$\hat{H}_b$,
$G(u)=\Tr \left[e^{iu\hat{H}_{\mathrm{b}}} \hat{U}(t,t_0)
  e^{-iu\hat{H}_{\mathrm{b}}} \hat{\rho}(t_0)
  \hat{U}^\dagger(t,t_0)\right],$ where $\hat{\rho}$ is the full
density matrix and $\hat{U}$ the full time evolution operator. $G(u)$
is the characteristic function -- the Fourier transform -- of the heat
distribution $P(Q)$. To compute it, one introduces a modified reduced
density matrix $\hat{\rho}_u$, such that $G(u)=\Tr
\hat{\rho}_u$. $\hat{\rho}_u$ evolves according to a modified
Hamiltonian, which includes phase factors that serve to keep track of
the transfer of energy, or indeed other quantities, to and from the
environment. These phase factors appear in the equation-of-motion\cite{murphy_quantum_2019} for $\hat{\rho}_u$, which is otherwise identical to Eq. (\ref{eq:diss}). They introduce a factor of $e^{-iu\Lambda(t)}$ multiplying the final term in
Eq. (\ref{eq:linsupop}) when it is used with the jump operator
$|+\rangle\langle-|$ and, similarly, a factor $e^{iu\Lambda(t)}$ when
it is used with the jump operator $|-\rangle\langle+|$. A consequence of this equation is that the mean heat current from the dot
to the phonons is given by \begin{equation}
  \frac{d\langle Q\rangle}{dt}=\Lambda(t)(\gamma_e p_+-\gamma_a
  p_-), \label{eq:heatcurr}\end{equation} where $p_{\pm}$ are the occupations of the
dressed states.

\section{Thermodynamic processes with chirped laser pulses}

\begin{figure}
\includegraphics{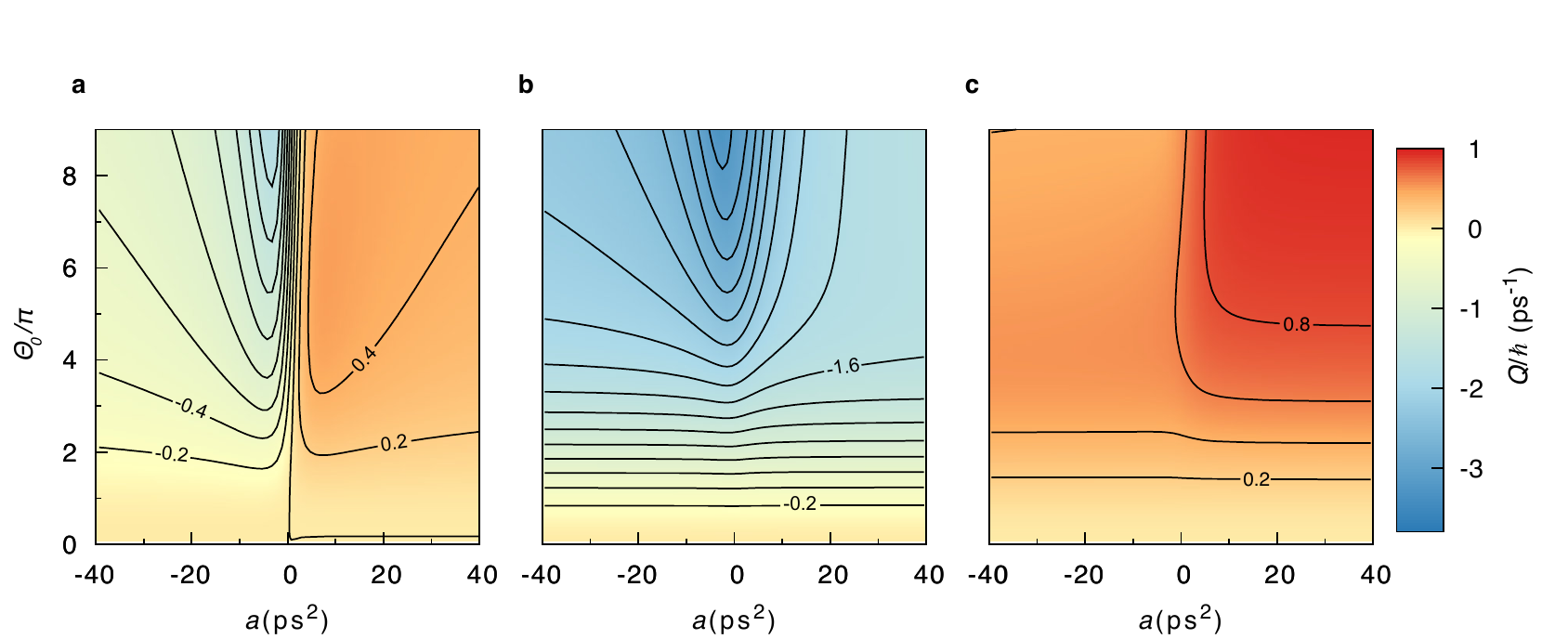}
\caption{Predicted heat transfer from the phonons to a quantum-dot exciton, with linearly-chirped driving pulses. The pulses are derived from a bandwidth-limited Gaussian pulse of duration $\tau_0=2\;\mathrm{ps}$, by applying a spectral chirp $a$ (horizontal axes). The different panels correspond to pulses whose center frequency is resonant with the transition (a), blue detuned from the transition, $\delta=-2.5\;\mathrm{ps}$ (b), and red detuned, $\delta=+2.5\;\mathrm{ps}$ (c). From Ref.\ \citenum{murphy_quantum_2019}.\label{fig:heatmaps}}
\end{figure}

We now present some results for the thermodynamic effects of driving a quantum-dot exciton with a chirped laser pulse, similar to that used to implement adiabatic rapid passage\cite{wu_population_2011,simon_robust_2011}.  Specifically, we consider linearly-chirped Gaussian pulses\cite{malinovsky_general_2001}, produced by introducing a spectral chirp $a$ to a bandwidth-limited pulse of duration $\tau_0$. We characterize the strength of the driving by the pulse area $\Theta_0$ of this bandwidth-limited pulse. Another relevant parameter is the detuning between the center frequency of the pulse and the exciton transition, $\delta=E_x-\omega_0$. Note that pulses with positive values of $\delta$ are red detuned with respect to the exciton. In all cases we assume a phonon bath temperature $T=20\;\mathrm{K}$.

\subsection{Heat absorption and emission}

Figure\ \ref{fig:heatmaps} shows the calculated mean heat transfer, from the phonons to the quantum-dot exciton, under the action of a linearly-chirped Gaussian pulse. We use this sign convention for heat, which differs from that used above by a sign, in the remainder of the paper. With it, positive values of $Q$ correspond to heat being transferred from the phonons to the dot, cooling the phonon environment. For the pulse with zero detuning, this occurs for positive chirp, while negative chirp produces the opposite effect. Detuning the pulse so that its spectrum lies above the exciton transition produces only heating of the phonons over the parameter regions shown, while detuning it below the exciton transition produces only cooling. 

These results can be understood from several perspectives, but they all come down to considering the time evolution of the populations and energies of the dressed-state levels, and the associated phonon emission and absorption processes. For example, we can explain the absorption of heat by a positively-chirped pulse, in the case of zero detuning, by noting that for positive chirp the initial state, $|0\rangle$, is the lower-energy dressed state, $|-\rangle$. There is no phonon emission from this state, so the system can evolve coherently, or absorb a single phonon and transition to the $|+\rangle$ state. If it does the latter then, as there is no phonon absorption from the $|+\rangle$ state, it can subsequently either evolve coherently, or emit a single phonon and return to the $|-\rangle$ state. For a trajectory involving absorption, and then emission, the heat transfers will cancel out to some extent. Thus the different possibilities for switching between the levels leads to a net absorption of heat, due to the asymmetry in the initial populations of the dressed states. Indeed, in this case, the initial populations correspond to a thermal distribution at zero temperature -- only the lowest-energy state is populated. Heat therefore flows from the phonons, at a non-zero temperature, to the dot. 

\subsection{Reversibility of heat transfers}

An important aspect of a thermodynamic process is the extent to which it is reversible, because reversible processes are those which produce maximum efficiency in a thermal machine such as a heat engine\cite{dann_quantum_2020}. The degree of reversibility of a process is quantified by the entropy it generates, taking into account the system and the heat baths. For an ideal reversible heat transfer, the entropy change of the system and heat bath are equal and opposite, and the entropy generation is zero. 

\begin{figure}
    \centering
    \includegraphics[width=6in]{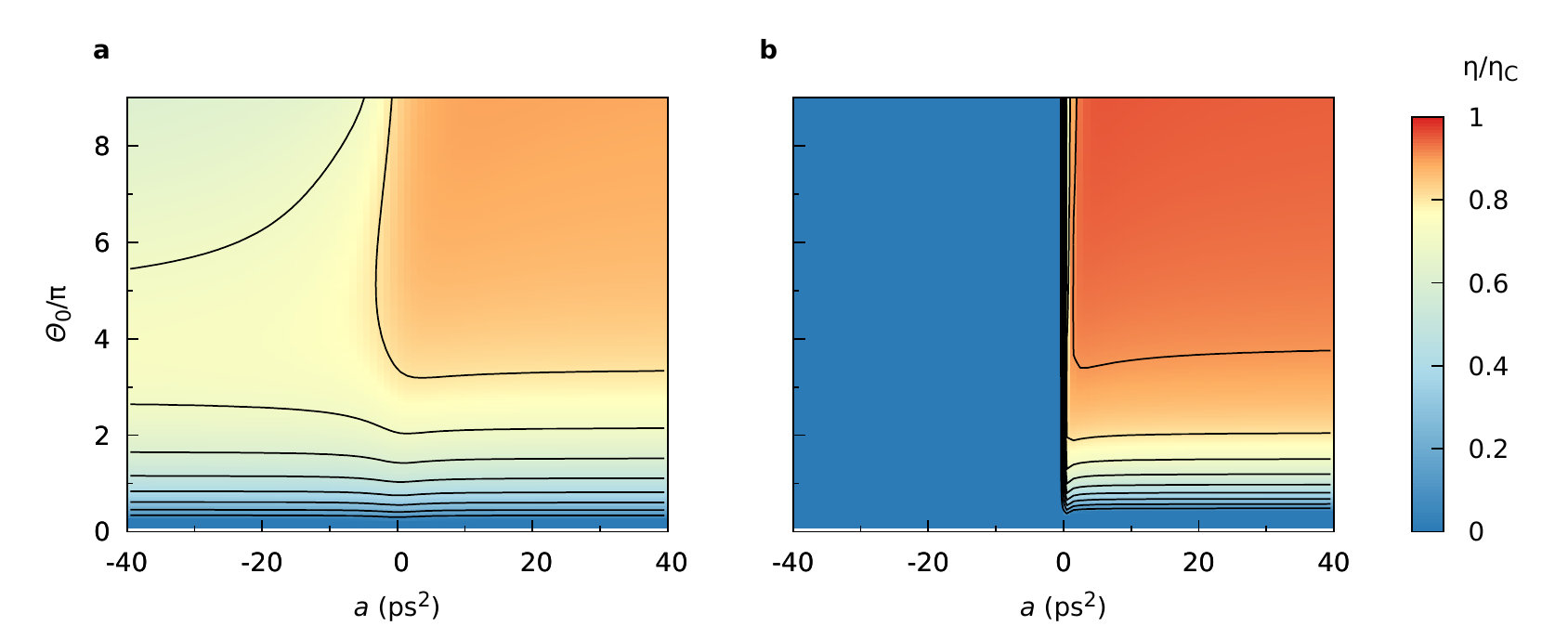}
    \caption{Predicted efficiency of a heat engine, in which the hot stroke is formed by driving a quantum-dot exciton with a linearly-chirped laser pulse. This causes it to absorb heat from the phonon bath at temperature $T_h=20\;\mathrm{K}$. The cycle is assumed to be closed by a reversible process, with a lower-temperature heat bath $T_c=2.7\;\mathrm{K}$. The efficiency is shown as a fraction of the Carnot efficiency $\eta_C$ for these temperatures. The detuning for both panels $\delta=2.5\;\mathrm{ps}^{-1}$. The pulse is assumed to be generated from a bandwidth-limited Gaussian of duration $\tau_0=2\;\mathrm{ps}$ (a), and $\tau_0=0.5\;\mathrm{ps}$ (b). Adapted from Ref.\ \citenum{murphy_quantum_2019}. }
    \label{fig:efficiency}
\end{figure}

A related, and perhaps more intuitive, way to quantify the reversibility of a process is to consider it as one stroke in a thermodynamic cycle. Figure.\ \ref{fig:efficiency} shows our predictions for the efficiency of a heat engine, in which one heat transfer stroke is provided by driving a quantum-dot with a laser pulse, inducing it to absorb phonons from a hot bath at $T_h=20\;\mathrm{K}$. We then suppose that, after this heat-absorption stroke, the cycle is closed by a reversible process, which transfers heat to a cold bath at a lower temperature $T_c<T_h$, does work, and returns the quantum-dot to its initial state. This forms a heat engine, operating between temperatures $T_h$ and $T_c$. The maximum efficiency is given by the Carnot efficiency $\eta_C=1-T_c/T_h$, and will be achieved when the entire cycle is reversible. The reduction in efficiency of our cycle, compared with $\eta_C$, reflects the irreversibility of the exciton-phonon heat transfer.

\begin{figure}
    \centering
\includegraphics{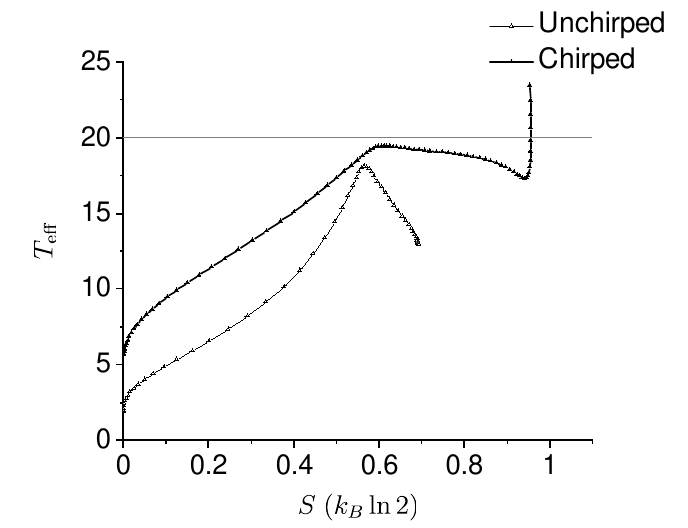}
    \caption{Predicted temperature and entropy of a quantum-dot exciton, for driving with a chirped, and an unchirped, laser pulse. The horizontal line shows the phonon bath temperature.}
    \label{fig:tsdiag}
\end{figure}

A notable feature of Fig.\ \ref{fig:efficiency} is that introducing positive chirp increases the efficiency, up to values of around $0.95\eta_C$. The physics behind this can be understood by plotting the thermodynamic state of the dot, during the driving pulse, on a temperature-entropy diagram, as shown in Fig.\ \ref{fig:tsdiag}. Here we compare two pulses: one with $\tau_0=2\;\mathrm{ps}$, $\Theta_0=6\pi$, and no chirp, and one with $\tau_0=0.5\;\mathrm{ps}$, $\Theta_0=9\pi$, and a chirp $a=10\;\mathrm{ps}^2$. The dressed-state energy levels for the chirped pulse are those shown in Fig.\ \ref{fig:carnotpulselevels}. The effective temperature of the dot is defined by noting that the gap between the dressed states is $\Lambda$, so we may equate the ratio of their populations to the Boltzmann factor $p_+/p_-=e^{-\Lambda/k_B T_{\mathrm{eff}}}$. For both pulses, the dot state starts at zero temperature, and heats up as phonons are absorbed. For the unchirped pulse the dot temperature is always significantly different to that of the phonon bath, so that the heat transfer is irreversible. For the chirped pulse, however, there is a region where the dot temperature is approximately constant, and slightly below that of the phonon bath. The heat transfer during this part of the pulse is an approximately isothermal, quasi-reversible process. It arises because the level splitting is reducing during this part of the pulse, as shown in Fig.\ \ref{fig:carnotpulselevels}. This drives the temperature of the dot down, and compensates for the heating that would otherwise occur as phonons are absorbed. 

\subsection{Heat distributions}

\begin{figure}
    \centering
    \includegraphics{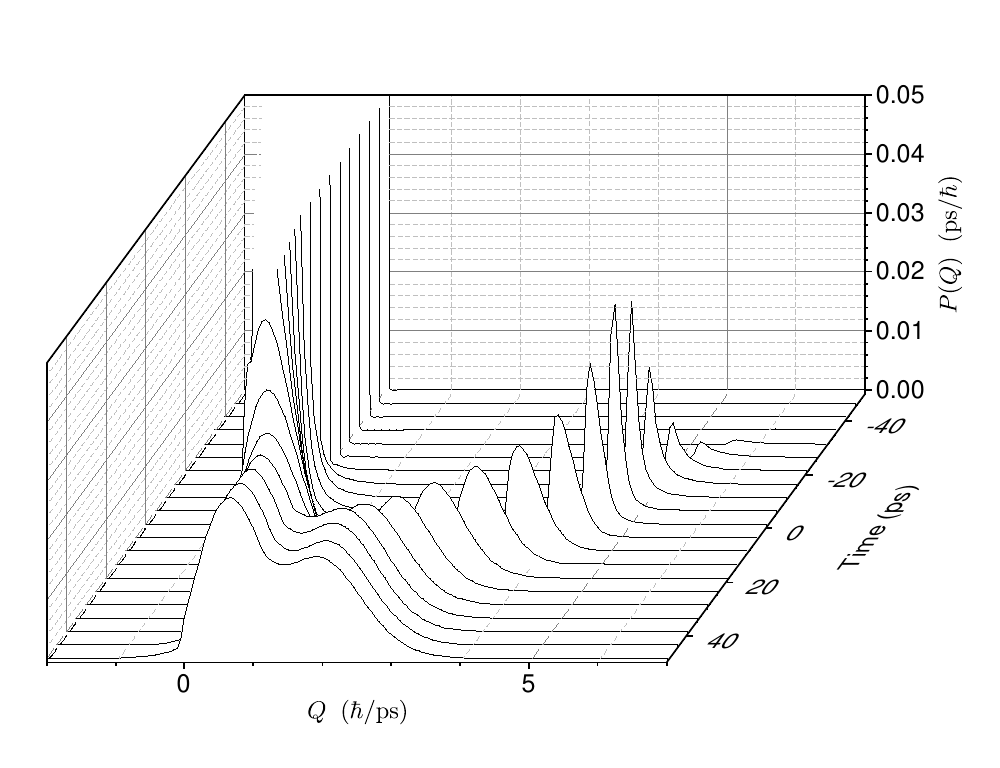}
    \caption{Probability distribution of the heat transferred from the phonons to the exciton, when the latter is driven with a chirped laser pulse.}
    \label{fig:heatdistn}
\end{figure}

One of the ways in which quantum thermodynamics differs from classical thermodynamics is that the heat is a stochastic quantity, with a probability distribution $P(Q)$. Figure\ \ref{fig:heatdistn} shows the heat distribution, for the chirped pulse that gives near-reversible heat transfer. This distribution is calculated by computing the characteristic function $G(u)$, as described in Sec.\ \ref{sec:modelling}, at discrete values of $u$, and Fourier transforming the result. 

This heat distribution can be understood in terms of the dressed-state energies for the pulse. Before the pulse arrives the heat distribution is simply a delta peak at $Q=0$, as there is no exchange of heat. This peak is artificially broadened by the finite sampling in $u$, but nonetheless exceeds the range of the scale. At the beginning of the pulse, another (broad) peak emerges at a high $Q$, corresponding to phonon absorption in the region of large dressed-state splitting. This peak moves to lower energies, and broadens, as phonons are exchanged with the bath over the reducing dressed state splitting. In addition, a shoulder develops and broadens on the delta peak at $Q=0$; this is due to processes where absorption of a phonon at one energy is followed by emission of a phonon at a lower energy. All these processes remove weight from the delta peak which, due its artificial broadening, disappears into the continuous features around the middle of the pulse. 

An interesting feature of the final heat distribution is the presence of a weak shoulder at negative $Q$. This means that, even though the process on average transfers heat from the phonons to the exciton, heat occasionally flows in the opposite direction. The effect arises from processes where a phonon is absorbed near the minimum splitting, and then re-emitted with a higher energy later in the pulse. Because of the choice of pulse parameters this is only a weak effect, but can be much more pronounced for other values of the parameters. The presence of fluctuations, with the opposite sign of heat transfer to the mean, emphasizes the stochastic nature of heat transfer in quantum systems\cite{miller_quantum_2020}. 

\section{Phonon cooling using dressed-state transitions}

\begin{figure}\centering
  \includegraphics{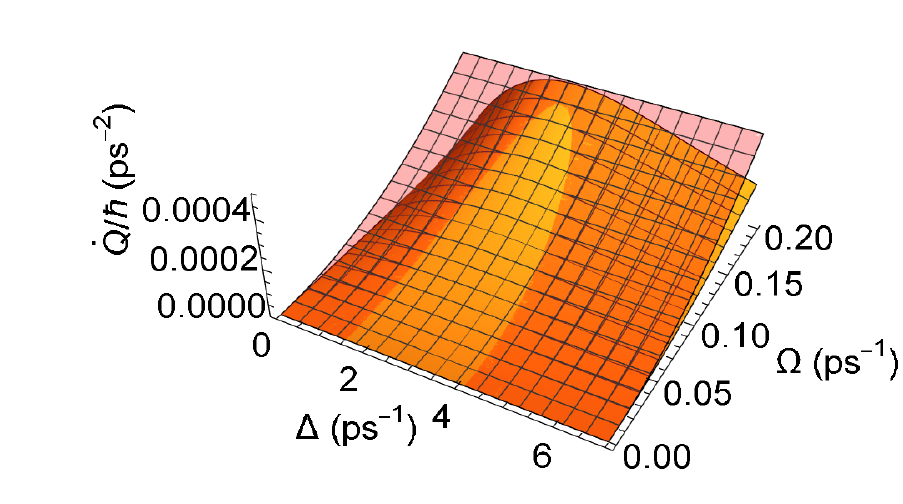}
  \caption{Yellow surface: Calculated cooling power of a silicon vacancy in diamond, driven by continuous-wave excitation with detuning $\Delta$ and Rabi splitting $\Omega$. The vacancy is modelled as a two-level system, in contact with a phonon bath at $T=20\;\mathrm{K}$. Pink surface: Heating per vacancy due to absorption of the driving laser, for $\rho/\alpha=1.47\times10^{22} \mathrm{m}^{-2}$ and a transition dipole $d=14.3\;\mathrm{Debye}$.}
  \label{fig:sivcool}
  \end{figure}

One potential application of this physics would be to implement a form of laser cooling, to remove heat from the acoustic phonons in the host crystal. This could perhaps be done using a cyclic process, which begins by driving the exciton transition, as above, with suitable pulse. The spontaneous decay of the exciton, which occurs over much longer timescales than those of the driving pulse, can then transfer the absorbed heat to the electromagnetic field. It will also return the dot to its ground state, allowing it to absorb heat again. 

Another possibility is to drive with continuous-wave excitation on the red side of the transition. The cooling effect of such driving, in conjunction with spontaneous emission, has been predicted by Gauger and Wabnig\ \cite{gauger_heat_2010}. It works because, for red detuning, the one-exciton state predominates in the upper dressed state, and the zero-exciton state in the lower. Spontaneous decay of the exciton thus preferentially depopulates the upper dressed state, in turn imbalancing the phonon transitions so they preferentially depopulate the lower dressed state, i.e., towards absorption.

However, the cooling power of a single quantum emitter is very low. The maximum heat such an emitter can absorb per spontaneous emission event is on the order of $k_B T$, so the cooling rate is at most $k_B T \gamma_{sp}$. Achieving a practically useful level of cooling therefore requires an ensemble of emitters, distributed through a host material. Unfortunately, for quantum-dot excitons the inhomogeneous broadening would seem to preclude achieving net cooling in an ensemble, at least using the steady-state protocol. The differing energies of the different transitions (due to variations in the dot size), mean that a detuning which cools one dot will lead to heating of another. The size of the quantum-dot, which comprises many thousands of atoms, is also problematical, because it means one cannot reach a high density of emitters. 

These considerations suggest that optically-active, lattice-scale defects are a better prospect for cooling than excitons in quantum-dots. We have considered, in particular, the possibility of cooling using the silicon vacancy center in diamond\cite{jahnke_electronphonon_2015}. Figure \ref{fig:sivcool} shows the cooling power per defect, for a two-level system model, Eq. (\ref{eq:tls}), using the phonon spectral density for a silicon vacancy\ \cite{norambuena_microscopic_2016}. This power is calculated by adding, to Eq. (\ref{eq:diss}), the term describing spontaneous emission, $\gamma_{sp} \mathcal{L}(|0\rangle\langle1|)$. We take\ \cite{jahnke_electronphonon_2015} $\gamma_{sp}=1\;\mathrm{ns}^{-1}$. We solve the resulting equations to determine the populations $p_{\pm}$ in the steady-state, and evaluate the heat current from Eq. (\ref{eq:heatcurr}). Note that the silicon vacancy is not, in fact, a two-level system: it has four orbital levels, comprising two spin-orbit split doublets separated by the optical frequency\ \cite{jahnke_electronphonon_2015}. The results in Fig.\ \ref{fig:sivcool} are therefore only indicative of the general effect.

The cooling effect of the vacancy will be offset by heating due to absorption, in the diamond host, of the driving laser. The power absorption per vacancy varies linearly with the driving intensity, and so quadratically with the Rabi frequency $\Omega$. It is, also, proportional to $\alpha/\rho$, where $\alpha$ is the absorption coefficient and $\rho$ the vacancy density. The heating effect is shown as the pink surface in Fig.\ \ref{fig:sivcool}, for an assumed value of $\rho/\alpha$. The existence of a region of net cooling can be seen. Note that for an absorption coefficient $\alpha=0.1\;\mathrm{cm}^{-1}$ the value of $\rho/\alpha$ assumed corresponds to an emitter density of $10^{23}\; \mathrm{m}^{-3}$, which is on the order of one silicon vacancy per million carbon atoms.

\acknowledgments 
 
We acknowledge funding from the Irish Research Council under award No. GOIPG/2017/1091, and Science Foundation Ireland under award No. 15/IACA/3402.


\end{document}